\documentclass[aps,prb,twocolumn,showpacs,superscriptaddress]{revtex4-1}
\usepackage{graphicx}
\usepackage{xcolor}   
\usepackage{amsmath}
\begin{document} 
\title{Thermoelectric transport through a quantum nanoelectromechanical system and its backaction}
\author{Hangbo Zhou} 
\affiliation{Department of Physics and Centre for Computational Science and Engineering, National University of Singapore, Repulic of Singapore 117551}
\affiliation{NUS Graduate School for Integrative Sciences and Engineering, National University of Singapore, Republic of Singapore 117456}
\author{Juzar Thingna}
\email[]{juzar.thingna@physik.uni-augsburg.de}
\affiliation{Institute of Physics, University of Augsburg, Universit\"atsstrasse 1 D-86135 Augsburg, Germany}
\affiliation{Nanosystems Initiative Munich, Schellingrstrasse 4, D-80799 M\"unchen, Germany}
\author{Jian-Sheng Wang}
\affiliation{Department of Physics and Centre for Computational Science and Engineering, National University of Singapore, Repulic of Singapore 117551}
\author{Baowen Li}
\affiliation{Department of Physics and Centre for Computational Science and Engineering, National University of Singapore, Repulic of Singapore 117551}
\affiliation{NUS Graduate School for Integrative Sciences and Engineering, National University of Singapore, Republic of Singapore 117456}
\affiliation{Centre for Phononics and Thermal Energy Science, School of Physics Science and Engineering, Tongji University, 200092 Shanghai, China}
\date{\today}

\begin{abstract}
We present a comprehensive study of thermoelectric transport properties of a quantum nanoelectromechanical system (NEMS) described by a single-electron-transistor (SET) coupled to a quantum nanomechanical resonator (NR). The effects of a quantum NR on the electronic current are investigated with special emphasis on how the SET-NR coupling strength plays a role in such a NEMS. We find that the SET-NR coupling is not only able to suppress or enhance the thermoelectric current but can also switch its direction. The effect of the NR on the thermoelectric coefficients of the SET is studied and we find that even a small SET-NR coupling could dramatically suppress the figure of merit $ZT$. Lastly, we investigate the backaction of electronic current on the NR and possible routes of heating or cooling the NR are discussed. We find that by appropriately tuning the gate voltage the backaction can be eliminated, which could find possible applications to enhance the sensitivity of detection devices.
\end{abstract}
\pacs{85.35.Gv, 85.85.+j, 05.60.Gg, 85.80.Fi}
\maketitle

\section{Introduction}
\label{introduction}
Nanomechanical resonators (NRs) have been in the limelight because of their possible applications in ultra-sensitive detection \cite{Kolkowitz2012, LaHaye2004, Mozyrsky2004} and quantum-controlled devices \cite{Benyamini2014, Eom2011}. Recent advances in high-frequency NR fabrication and cooling technology have made it possible to achieve quantum behaved resonators \cite{Santandrea2011, Naik2006}, which opens up the possibility for mechanical systems to be coherently coupled to electronic ones to form a quantum nanoelectromechanical system (NEMS) \cite{Gaidarzhy2005, Li2012a, Naik2006,Piovano2011}. Such NEMSs have shown a wide range of applications such as charge probing \cite{Meerwaldt2012}, coherent sensing\cite{Kolkowitz2012, Palyi2012}, and electron shuttling \cite{Moskalenko2009}. In such applications an important parameter of the NEMS is the coupling strength between the NR and the electronic system. Traditionally, the coupling strength is assumed to be weak \cite{Rodrigues2005,Blanter2004,Armour2004}, hence the effect of the NR on the electronic system is often regarded as a perturbation. However, recent experiments have demonstrated that the coupling between the NR and electronic system can be strong \cite{Steele2009, Lassagne2009} and even be tailored \cite{Benyamini2014, Lassagne2009}. The influence of the NR on the electronic system in this strong coupling regime is a facet which has not yet been fully explored theoretically under a quantum mechanical description.

An important class of the NEMSs is an NR coupled to a single-electron-transistor (SET-NR system), which has been widely investigated both theoretically \cite{Clerk2005,Blanter2004,Armour2004} and experimentally \cite{Mozyrsky2004, LaHaye2004, Lassagne2009}. In most cases, the NR is treated classically under the condition that the resonant frequency is much smaller than the electron tunneling frequency. The transport properties of such classical SET-NR systems have been extensively studied, including the calculation of current, current noise \cite{Harvey2008, Armour2004}, and dynamics of the NR \cite{Kirton2013, Nocera2012, Blencowe2005, Armour2004}. Among these investigations an interesting perspective is to understand how the NR affects the electronic current\cite{Nocera2012, Kirton2013,Blencowe2005,Wang2012}, and hence facilitates the analysis of mechanically tuned electronic signals from the detectors, or helps in the design of mechanically controlled electronic nano-devices. Previous studies have shown that the NR causes changes to electronic current near the resonant frequency\cite{Nocera2012} and this phenomenon is widely used in experiments\cite{Benyamini2014,Lassagne2009,Steele2009} for various applications.  However, these works focus on the influence of NR on the voltage biased current. The influence of a quantum NR on thermoelectric current, where the electronic current is induced by a temperature bias, is largely ignored. Furthermore, when the resonant frequency of NR is comparable to or even larger than the electron tunneling frequency, the SET-NR coupling exhibits quantum behavior and the coherences between the NR and the SET become important \cite{Piovano2011}. Theoretical investigations taking the coherence between NR and SET are quite limited. In this work we study transport properties of a complete quantum mechanical SET-NR system and find intriguing effects due to the influence of the NR on thermoelectric current. In particular, the NR is not only able to adjust the magnitude of current as previously discovered, but also able to tune its direction which is impossible in voltage biased mode.

Another important aspect to fully understand the transport properties is the backaction on the NR caused by the passage of electronic current. Experimentally the backaction has been constructively employed to cool the NR to the quantum regime\cite{Clerk2005,Naik2006,Zippilli2010}. However, for most applications, such as displacement detection, the backaction is not preferred since it generates noise which in turn reduces the sensitivity of the detectors\cite{Mozyrsky2004,Mozyrsky2002}. For voltage biased current backaction is generally not avoidable and hence it imposes a fundamental limit to measurement sensitivity. However, in this work we find that by using thermoelectric current it is possible to fully eliminate the backaction.

In the following sections we present in detail how the thermoelectric current varies with the SET-NR coupling strength. We then demonstrate a complete picture of how a quantum NR affects the thermoelectric coefficients of the SET including the Seebeck coefficients and figure of merit. We then go on to study the backaction of the electronic system on the NR and the role of SET-NR coupling played in the backaction. Our comprehensive study allows us to fully understand the thermoelectric transport properties of the SET-NR system in the quantum regime and provides a better understanding of how to eliminate the backaction.

\section{Physical model and formalism}
\label{model}
\begin{figure}
\centering
\includegraphics[width=\columnwidth]{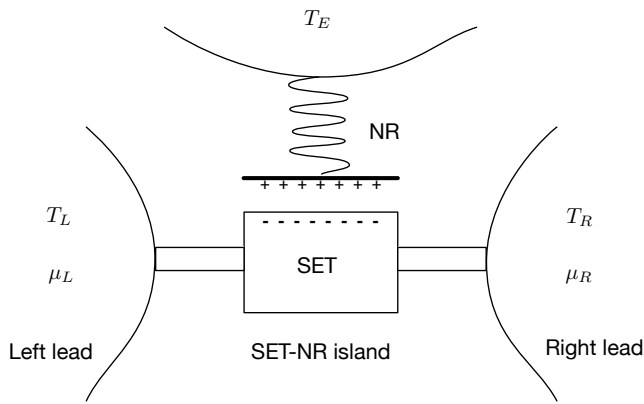}
\caption{\label{fig:schematic} A schematic of the SET-NR system studied in this work. The SET center is capacitively coupled to the NR to form a SET-NR island. The capacitance depends on the position of the NR. The SET-NR island is connected to two electronic leads with temperature $T_L$ ($T_R$) and chemical potential $\mu_L$ ($\mu_R$) for the left (right) lead. A background environment with temperature  $T_{E}$ acts on the NR.}
\end{figure}

We consider a SET-NR system where a SET is capacitively coupled to an NR as schematically shown in Fig.~\ref{fig:schematic}. The full setup consists of a SET-NR island coupled to two electrodes and the NR is also subject to a background environment with temperature $T_E$. The SET and NR are interacting through a gate capacitor with capacitance $C_g(x)$. The capacitor is built such that the capacitance depends on the displacement of the NR, $x$. In such nanostructures the voltage on the dot $V_{\mathrm{dot}}$ is sensitive to the excess charges on the dot due to the small capacitance of the SET-NR island, and it is governed by \cite{Steele2009, Lassagne2009,Meerwaldt2012}
\begin{equation}
V_{\mathrm{dot}}=\frac{-e}{C_\Sigma}(N-N_g),
\end{equation}
where $N$ is the number of excess charges on the dot, $e$ is the unit charge, $C_\Sigma$ is the total capacitance of the SET, and $N_g=C_gV_g/e$ denotes the control charge induced by the gate at voltage $V_g$. Hence the capacitance energy of the island is given by \cite{Nocera2012,Armour2004, Rodrigues2005, Makhlin2001}
\begin{equation}
E_N=E_C(N-N_g)^2,
\end{equation}
where $E_C=e^2/(2C_\Sigma)$ is the charging energy of the island. When the capacitance is so small that the charging energy dominates over the thermal fluctuations, the SET-NR island will be in the strict Coulomb blockade regime such that only a single electron can tunnel through the island. Therefore, the states of excess charge on the island are limited to two cases, $N$ or $N+1$ excess charge. We set the zero of energy at the energy level with $N$ excess charges. Hence, the energy of the system with $N+1$ excess  charges will be\cite{Blencowe2005}
\begin{equation}
\varepsilon=E_C(2N-2N_g+1).
\end{equation}
The energy $\varepsilon$ above depends on the displacement of the NR via $E_C$ and $N_g$. In experimental set-ups the displacement $x$ is much smaller than the separation distance between the SET and the NR. Hence we can expand the energy to linear order in $x$ as $\varepsilon=\varepsilon_0+\lambda x$, where $\varepsilon_0$ is a constant evaluated at equilibrium separation when there are $N$ excess charges. The coefficient $\lambda$ is given by
\begin{equation}
\lambda=\frac{-eC'_g}{C_\Sigma^2}\bigl[C_\Sigma V_g+(N-N_g+1/2)e\bigr].
\label{eq:lambda}
\end{equation}
Here $C'_g$ is the derivative of $C_g$ with respect to $x$ which is assumed to be a constant.  Equation~(\ref{eq:lambda}) is consistent with different forms obtained in the literature \cite{Meerwaldt2012,Steele2009,Lassagne2009} in the sense that the coefficient $\lambda$ above can be physically interpreted as the extra force acting on the NR by the SET when an excess charge is tunneling into the island. It also characterizes the coupling strength between the SET and NR and it is possible to be tuned by adjusting the oscillation mode of the NR\cite{Benyamini2014}. The constant $\varepsilon_0$ is proportional to the gate voltage $V_g$ and it can also be adjusted experimentally. Hence in the following discussion we will use $\varepsilon_0$ instead of $V_g$ to describe the gate properties. 

By writing the Hamiltonian in the basis of excess charges we can quantize the SET-NR island to obtain\cite{Makhlin2001}
\begin{equation}
H_{S}=(\varepsilon_0+\lambda x)|N\!+\!1\rangle\langle N\!+\!1| +H_{NR},
\end{equation}
where $x$ is a quantum mechanical position operator. The Hilbert space of the excess charges is spanned by $|N\rangle$ and $|N\!+\!1\rangle$ with fixed $N$. The NR Hamiltonian $H_{NR}$ is the standard harmonic oscillator 
\begin{equation}
H_{NR}=\frac{p^2}{2m}+\frac{1}{2}m\omega_0^2 x^2,
\end{equation}
where $\omega_0$ is the fundamental frequency of the NR and $m$ is its mass. The SET-NR island is connected to two electronic leads which act as a source and a drain. Since the experimental set-up cannot be isolated from dissipative effects acting on the NR, we subject the NR to an external environment which is kept at temperature $T_E$. Therefore, the total Hamiltonian reads
\begin{equation}
H_{tot}=H_S+H_{L}+H_{R}+H_{E}+H_T+H_{SE},
\end{equation}
where the lead Hamiltonians $H_{L,R}$ are modeled as an infinite set of free-fermions given by
\begin{equation}
H_{L,R} = \sum_{k\in L,R}\epsilon_kc^\dagger_kc_k.
\end{equation}
Above $c_k^{\dagger}$ and $c_k$ are the fermionic creation and annihilation operators respectively. The environment Hamiltonian induces dissipative effects and is assumed to be a phononic heat bath given by
\begin{eqnarray}
H_{E}=\sum_{n=1}^\infty\frac{p_n^2}{2m_n}+\frac{1}{2}m_n\omega_n^2 q_n^2.
\end{eqnarray}
The coupling of the system to this background environment is via the linear position coupling, $H_{SE}=-\sum_{n=1}^\infty C_n q_n x/\sqrt{2\omega_0}$. The tunneling Hamiltonian $H_T$ connects the SET and the electrodes which can be written as
\begin{equation}
H_T=\sum_{k\in L,R}V_k\left(c_k^\dagger |N\rangle\langle N\!+\!1|+c_k|N\!+\!1\rangle\langle N|\right),
 \end{equation}
where the first term describes the tunneling of a charge from the island to the lead while the second term expresses the reverse process. The information of the electrodes, environment and their coupling to the SET-NR island can be summarized using the spectral density $\Gamma(\varepsilon)$ (for the electronic leads) and $J(\omega)$ (for the environment) given by
\begin{eqnarray}
\Gamma_\alpha(\varepsilon)&=& 2\pi\sum_{k\in\alpha = 1}^{\infty}|V_k|^2\delta(\varepsilon-\varepsilon_k),\;\alpha=L,R \\
J(\omega) &=& \pi\sum_{n=1}^{\infty}\frac{c_n}{2m_n\omega_n}\delta(\omega-\omega_n).
\end{eqnarray}
In the thermodynamic limit of the leads and the environment the spectral densities become a continuous function and in this work we choose them to be of the form
\begin{eqnarray}
\label{eq:spece}
\Gamma_\alpha(\varepsilon)&=&\frac{\eta_\alpha}{1+(\varepsilon/\varepsilon_D)^2},\;\alpha=L,R \\
\label{eq:spech}
J(\omega)&=&\frac{\eta_{E}\omega}{1+(\omega/\omega_D)^2},
\end{eqnarray}
where $\eta_{L},\eta_R$ (units of energy), $\eta_E$ (units of $\hbar$) represent the square of the system-lead and system-environment coupling strength and $\varepsilon_{D}$ ($\omega_{D}$) represents a cut-off energy (frequency) of the Lorentz-Drude form to avoid ultra-violet divergences \cite{Dittrich1998}. 

The Hamiltonian described above has been well-discussed in the literature, mostly in the study of vibrational effects in molecular junction systems \cite{Chen2005,Sun2007, Zhou2012, Koch2005, Arrachea2014, Koch2011,Koch2014, Zianni2010, Piovano2011,Perroni2014, Leijnse2010,Lu2007,McEniry2008a,Asai2008,Lee2009}. In molecular junctions the phonon arises either due to the center-of-mass motion of the molecule \cite{Park2000} or it could be thermally induced \cite{Qin2001} thus limiting the number of phonons interacting with the electronic degrees of freedom. This differs significantly from our set-up due to the presence of a thermal environment which is weakly coupled to a single phonon in the system. The phonon environment emerges naturally in the context of SET-NR systems since the mechanical oscillator can not be isolated from the environment which damps its oscillatory motion.

In previous studies, polaron transformation has often been employed to decouple the electronic and vibrational degrees of freedom. However, in order to solve the problem one needs to introduce further approximations, which depend on the parameter regimes of interest. For example, a common assumption is to treat the vibrational mode at an equilibrium (canonical) distribution \cite{Chen2005,Sun2007,Zhou2012}, which is valid only when the SET-NR coupling is weak compared to the NR-environment coupling. Another approach is to use rate equations in the product basis of the SET and NR \cite{Koch2005,Kirton2013,Arrachea2014}, which essentially ignores the coherences between the SET and NR. Other treatments include the non-equilibrium Green's function technique\cite{Lu2007,McEniry2008a,Asai2008,Lee2009} for weak SET-NR coupling systems, or classical (semiclassical)\cite{Piovano2011,Perroni2014} treatments for slow vibration. However, we would like to study the system in the quantum regime of strong SET-NR coupling and hence we resort to the alternative of treating the system-lead and system-environment coupling as weak while the nonlinearity in SET-NR coupling is treated \emph{exactly}. Within this weak system-lead coupling approximation we employ the standard techniques of the theory of open quantum systems to write a quantum master equation of the Bloch-Redfield type given by\cite{Redfield1965, breuer2007theory}
\begin{eqnarray}
\label{eq:rho}
\frac{\partial\rho}{\partial t}&=&-\frac{\mathrm{i}}{\hbar}[H_S,\rho]-\frac{1}{\hbar^2}\sum_{\alpha ,\beta}\\
&&\times\int_{-\infty}^td\tau\bigl\{[S^\alpha,S^\beta(\tau-t)\rho]C^{\alpha\beta}(t-\tau)+\mathrm{H.c.}\bigr\}, \nonumber
\end{eqnarray}
where $\mathrm{H.c.}$ stands for Hermitian conjugate. Above $\rho$ is the reduced density matrix describing the state of the SET-NR island obtained by tracing over the lead and environment degrees of freedom and $S=\left\{|N\rangle\langle N\!+\!1|\,,\,|N\!+\!1\rangle\langle N|\,,\,x/\sqrt{2\omega_0}\right\}$ is a vector with each component denoting a particular system operator coupled to the leads and the environment. The corresponding operator of the leads or environment can also be expressed in a vector form and is given by, $B=\left\{\,\sum_{k\in L,R}V_kc_k^\dagger\,,\sum_{k\in L,R}V_kc_k\,,\,-\sum_{n=1}^\infty C_n q_n\right\}$. Subsequently the correlation functions used in Eq.~(\ref{eq:rho}) are defined as $C^{\alpha\beta}(t)=\langle B^\alpha(t)B^\beta(0)\rangle$, where the superscript denotes a particular component of the $B$-vector. Operators with time arguments indicate free evolutions with respect to the non-interacting Hamiltonian $H_0=H_S+H_{L}+H_{R}+H_{E}$. The non-vanishing correlation functions defined above can be expressed as
\begin{eqnarray}
C^{12}(t)&=&\sum_{\alpha = L,R}\int_{-\infty}^\infty \frac{d\epsilon}{2\pi}\Gamma_\alpha(\epsilon)f_\alpha(\epsilon)e^{\mathrm{i}\epsilon t/\hbar},\\
C^{21}(t)&=&\sum_{\alpha = L,R}\int_{-\infty}^\infty \frac{d\epsilon}{2\pi}\Gamma_\alpha(\epsilon)\bigl[1-f_\alpha(\epsilon)\big]e^{-\mathrm{i}\epsilon t/\hbar}\!\!, \\
C^{33}(t)&=&\int_{-\infty}^{\infty}\frac{d\omega}{\pi}J(\omega)n(\omega)e^{\mathrm{i}\omega t},
\end{eqnarray} 
where $f_\alpha(\epsilon)=\left[e^{\beta_\alpha(\epsilon-\mu_\alpha)}+1\right]^{-1}$is the Fermi-Dirac distribution of the $\alpha$th lead, $n(\omega)=\left[e^{\beta_E\hbar\omega}-1\right]^{-1}$ is the Bose-Einstein distribution and we have assumed that the left and right leads are uncorrelated. In order to obtain the correlation function of the environment we have assumed $J(-\omega) = - J(\omega)$, which is true for ohmic spectral density chosen in this work. Above $\beta_\alpha=\left[k_BT_\alpha\right]^{-1}$ and $\mu_\alpha$ correspond to the inverse temperature and chemical potential of the leads and the environment with appropriate subscripts $\alpha$. The analytical forms of these correlation functions can be found in the Appendix \ref{sec:correlation}.

In order to solve Eq.~(\ref{eq:rho}) we numerically diagonalize the system Hamiltonian $H_S$ so that the SET-NR coupling is treated exactly. Thus, in the energy eigenbasis of the system Hamiltonian $H_{S}$ the quantum master equation reads
\begin{equation}
\frac{d\rho_{nm}}{d t}=-\frac{\mathrm{i}}{\hbar}\Delta_{nm}\rho_{nm}+\sum_{ij}R^{ij}_{nm}\rho_{ij},
\end{equation}
where
\begin{eqnarray}
R^{ij}_{nm}&=&\frac{1}{\hbar^2}\sum_{\alpha ,\beta}\Bigl\{ S^\alpha_{ni}S^\beta_{jm}W_{ni}^{\alpha\beta}\nonumber\\
                    &&-\delta_{jm}\sum_lS^\alpha_{nl}S^\beta_{li}W_{li}^{\alpha\beta}\Bigr\}+\mathrm{H.c.}
\end{eqnarray}
The transition coefficients are given by
\begin{equation}
W_{ij}^{\alpha\beta}=\int_{-\infty}^{t} d\tau e^{\mathrm{i}\Delta_{ij}(\tau-t)/\hbar}C^{\alpha\beta}(t-\tau),
\end{equation}
where $\Delta_{ij}=E_i-E_j$ is the energy spacing of the system Hamiltonian.

Since we are interested in the steady-state thermoelectric transport properties we solve the above quantum master equation in the steady state by setting $d\rho/dt=0$ and taking the limit $t \rightarrow 0$. In order to evaluate the currents at the lowest order we require only the $0$th order reduced density matrix\footnote{The $0$th order reduced density matrix is obtained by expanding the density matrix $\rho = \rho^{(0)}+\mathcal{G}(V_{k}^{2},C_{n}^{2})\rho^{(2)}+\cdots$, where $\mathcal{G}$ is a function of the system-lead and system-environment coupling strength squared.}, which is obtained by solving $\sum_{i}R^{ii}_{nn}\rho^{(0)}_{ii}=0$ along with the normalization condition $\mathrm{Tr}(\rho^{(0)})=1$. The off-diagonal elements satisfy $\rho^{(0)}_{ij} = 0\, \forall\, i \neq j$. We would like to point out that $\rho^{(0)}$ is diagonal in the eigenbasis of the system Hamiltonian $H_S$ and hence it will have off-diagonal elements in the product basis of SET and NR, which implies that the coherences of the SET-NR are properly taken into account. The price to pay in order to use only the $0$th order reduced density matrix \cite{Thingna2014, Juzar2012} is that we then require the information about the system-lead coupling in the current operator. This obstacle can be overcome if we follow the techniques of Thingna \emph{et~al.} in Refs.~[\onlinecite{Juzar2012,Wang2013review, Thingna2014}] to obtain the reduced definition of the current operator. This reduced current operator could then be combined with the $0$th order reduced density matrix $\rho^{(0)}$ to obtain the average currents at the lowest order of system-lead coupling. Thus, in order to obtain the reduced definition of the current operators we begin with the standard definition \cite{Meir1992} of the electron and heat current operators out of the left lead as
\begin{eqnarray}
\mathcal{I}_e&=&-e\frac{dN_L}{dt} \nonumber \\
&=&\frac{\mathrm{i}e}{\hbar}\sum_{k\in L}V_k\left(c_k^\dagger |N\rangle\langle N\!+\!1|-c_k|N\!+\!1\rangle\langle N|\right), \\
\mathcal{I}_{h}&=&-\frac{d (H_L-\mu_LN_L)}{dt}\nonumber \\
&=&\frac{\mathrm{i}}{\hbar}\sum_{k\in L}V_k(\varepsilon_k-\mu_L)\left(c_k^\dagger|N\rangle\langle N\!+\!1|-c_k|N\!+\!1\rangle\langle N|\right),\nonumber\\
\end{eqnarray}
where $\mu_L$ represents the chemical potential of the left lead and $N_L=\sum_{k\in L}c^\dagger_kc_k$ is the left lead electron number operator.

Instead of treating the electron and heat current separately we use a unified notation,
\begin{eqnarray}
\label{eq:curr}
\mathcal{I}_{e(h)}=\frac{\mathrm{i}}{\hbar}\sum_{\alpha=1,2}S^\alpha\otimes \mathcal{B}_{e(h)}^\alpha,
\end{eqnarray}
to treat both currents on the same footing. Above the operator-vectors, $\mathcal{B}_e=\left\{\,e\sum_{k\in L}V_kc_k^\dagger,-e\sum_{k\in L}V_kc_k\,\right\}$ and $\mathcal{B}_h=\left\{\,\sum_{k\in L}(\varepsilon_k-\mu_L) V_kc_k^\dagger,-\sum_{k\in L}(\varepsilon_k-\mu_L) V_kc_k\,\right\}$, contain information about the left lead. Following the derivation of Refs.~[\onlinecite{Juzar2012,Wang2013review, Thingna2014}] we obtain the lowest order of the reduced steady-state current operators as
\begin{eqnarray}
\mathcal{I}^r_{e(h)}&=&\frac{1}{\hbar^2}\sum_{\alpha,\beta}\int_{-\infty}^{0}d\tau S^\alpha S^\beta(\tau)\mathcal{C}_{e(h)}^{\alpha\beta}(-\tau)+\mathrm{H.c.},
\end{eqnarray}
where $\mathcal{C}_{e(h)}^{\alpha\beta}(\tau)=\langle B^\alpha(\tau)\mathcal{B}_{e(h)}^\beta(0)\rangle$ are the correlation functions between the lead operators occurring in the current operator definition of Eq.~(\ref{eq:curr}) and the tunneling Hamiltonian $H_{T}$. For the electron current operator the non-vanishing parts of these correlation functions are given by
\begin{eqnarray}
\mathcal{C}_e^{12}(t)&=&-e\int_{-\infty}^\infty \frac{d\epsilon}{2\pi}\Gamma_L(\epsilon)f_L(\epsilon)e^{\mathrm{i}\epsilon t/\hbar},\\
\mathcal{C}_e^{21}(t)&=&e\int_{-\infty}^\infty \frac{d\epsilon}{2\pi}\Gamma_L(\epsilon)\bigl[1-f_L(\epsilon)\big]e^{-\mathrm{i}\epsilon t/\hbar}.
\end{eqnarray}
 The non-vanishing correlation functions in the heat current operator are scaled by a factor of energy and can be expressed as
\begin{eqnarray}
\mathcal{C}_h^{12}(t)&=&\int_{-\infty}^\infty \frac{d\epsilon}{2\pi}(\epsilon-\mu_{L})\Gamma_L(\epsilon)f_L(\epsilon)e^{\mathrm{i}\epsilon t/\hbar},\\
\mathcal{C}_h^{21}(t)&=&\int_{-\infty}^\infty \frac{d\epsilon}{2\pi}(\epsilon-\mu_{L})\Gamma_L(\epsilon)\bigl[1-f_L(\epsilon)\big]e^{-\mathrm{i}\epsilon t/\hbar}.\nonumber\\
\end{eqnarray}
The analytical expressions for these correlation functions will be discussed in Appendix \ref{sec:correlation}.

Expressing the reduced current operators in the energy eigenbasis of $H_{S}$ we obtain
\begin{equation}
(\mathcal{I}^r_{e(h)})_{ij}=\frac{1}{\hbar^2}\sum_{\alpha,\beta, k}\Bigl[S^\alpha_{ik}S^\beta_{kj}\mathcal{W}^{\alpha\beta}_{e(h)}(\Delta_{kj})+c.c.\Bigr],
\end{equation}
where $c.c.$ denotes complex conjugate and the transition coefficients are different from that used in the quantum master equation and are given by
\begin{equation}
\mathcal{W}_{e(h)}^{\alpha\beta}(\Delta_{kj})=\int_{-\infty}^0 d\tau e^{i\Delta_{kj}\tau/\hbar}\mathcal{C}_{e(h)}^{\alpha\beta}(-\tau).
\end{equation}
Now since the reduced current operators are known, the average steady-state currents can be easily calculated using the $0$th order reduced density matrix as $I_{e(h)} = \mathrm{Tr}\left(\rho^{(0)}\mathcal{I}^r_{e(h)}\right)$. 

To explore the thermoelectric properties of the SET-NR system, we also need to calculate the transport coefficients in the linear response regime. In this regime the relation between heat and electron current can be expressed in a matrix form as
\begin{equation}
 \begin{pmatrix}I_e \\I_h \end{pmatrix} =\begin{pmatrix}L_0 & L_1 \\L_1 & L_2 \end{pmatrix}  \begin{pmatrix}\Delta\mu \\\Delta T/T \end{pmatrix}.
\end{equation}
Here, the off-diagonal elements $L_1$ are the same due to the Onsager reciprocal relations. Thus, we can express the transport coefficients in terms of the above matrix coefficients as\cite{Dubi2011}
\begin{eqnarray}
G_e &=& eL_0, \\
S &=& \frac{L_1}{G_eT}, \\
\kappa &=& \frac{L_2}{T}-G_eS^2T, \\
ZT &=& \frac{G_eS^2T}{\kappa},
\end{eqnarray}
where $G_e$ is the electronic conductance, $\kappa$ is the thermal conductance, $S$ is the Seebeck coefficient, and $ZT$ is the thermoelectric figure of merit which determines the efficiency of the device to convert (waste) thermal energy into (useful) electrical current.

\section{results and discussion}
\label{result}
\subsection{Thermoelectric transport}
\label{ZT}
\begin{figure}
\centering
\includegraphics[width=\columnwidth]{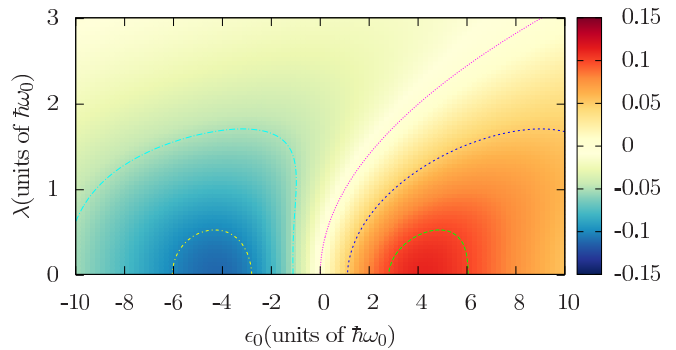}
\caption{\label{fig:current} {(Color online)} Electron current under a temperature bias in units of $e\eta/\hbar$. The background temperature is set at $T_{E}=5\hbar\omega_0/k_B$. The left and and right leads are at temperature $T_{L}=T_{E}+\Delta T$ and $T_{R}=T_{E}-\Delta T$ with $\Delta T=3\hbar\omega_0/k_B$. The couplings between the left and right leads are kept symmetric, $\eta=\eta_L=\eta_R$, while the coupling to the environment is fixed at $\eta_E=0.05\eta\omega_0$. The chemical potentials are set to $0$ for all leads.}
\end{figure}
The thermoelectric effects in vibrational coupled systems have been investigated mostly in the context of molecular junctions \cite{Leijnse2010, Koch2014, Perroni2014, Dubi2011, Ren2012, Zianni2010, Hsu2012}. Previous studies in such systems indicate that the vibrational effects on thermoelectric efficiency is very sensitive to the system and bath parameters, such as the SET energy level \cite{Koch2014, Perroni2014, Ren2012, Leijnse2010}, the coupling strength of the vibrational modes to its environment \cite{Leijnse2010}, the frequency of the vibrational mode \cite{Koch2014,Ren2012,Perroni2014}, the temperature \cite{Ren2012, Hsu2012}, and the chemical potential \cite{Leijnse2010, Zianni2010} of the system. However, both enhancement \cite{Ren2012,Koch2014} and suppression \cite{Zianni2010,Perroni2014} of the thermoelectric effects due to vibration were reported in different parameter regimes. In this work we systematically investigate the effect of SET-NR coupling on thermoelectric transport in all regimes of system parameters, with emphasis on the SET-NR coupling strength dependence. We provide an overall picture indicating the enhancement and suppression regimes for electronic conductance, thermal conductance, Seebeck coefficient, and the figure of merit $ZT$.

We first study the effect of NR on the thermoelectric current and its dependence on gate voltage, governed by $\varepsilon_0$, and SET-NR coupling strength $\lambda$ as shown in Fig.~\ref{fig:current}. From now onwards, we measure all quantities in units of $\hbar\omega_{0}$ and choose the parameters such that the thermal energy $k_B T$, chemical potential bias $e\Delta V$, and energy scale of the SET-NR island are comparable to $\hbar\omega_0$. This choice of parameters allows us to work in the quantum regime and is experimentally realizable since it is possible to create NRs with $\omega_0$ up to a few gigahertz \cite{Gaidarzhy2007}, which would correspond to the background temperature of tens of millikelvins \cite{Gaidarzhy2005} and a SET energy scale in the micro-eV range. From Fig.~\ref{fig:current} we clearly see that when $\lambda=0$ the thermoelectric current changes direction when $\varepsilon_0$ transverses across the Fermi-level of the leads. This can be easily understood from the fact that the dominant charge carriers of the SET are altered by varying the gate voltage\cite{Sierra2014}. Interestingly, we can also see that the thermoelectric current changes sign with the SET-NR coupling strength. This particularly happens for $\varepsilon_{0}$ larger than the Fermi-level of the leads. For example if we focus on $\varepsilon_{0}=\hbar\omega_0$ we find that at around $\lambda \approx 1.2\hbar\omega_0$ the current changes sign from positive to negative. The red dotted line on the contour plot helps locate the boundary of the sign change. This phenomenon is particularly interesting because it clearly demonstrates that the mechanical motion of the SET-NR structure is able to influence the dominant charge carriers inside the SET. One could understand this change of dominant charge carriers from electrons to holes in terms of a polaron shift process. When the quantized vibration of the NR couples to the SET it forms a polaron, which in turn shifts the energy spacing $\varepsilon_0$ downwards by an amount $\lambda^{2}/(\hbar\omega_0)$. Thus the stronger the SET-NR coupling strength $\lambda$, the larger would be the energy shift. Once the polaron shift becomes large enough to shift the energy from above the Fermi-level of the leads to below, we see a change in the dominant carrier type of the SET. In term of the magnitude of the current, in most regimes the SET-NR coupling will decrease the current. However, in some domains we also find that SET-NR coupling can be used to increase the current. For example, when $\varepsilon_0=-\hbar\omega_0$ the magnitude of current increases to a maximum at $\lambda=1.5\hbar\omega_0$ and then decreases again as a function of $\lambda$. The enhancement effect is even more pronounced in the low temperature regime. These results clearly exhibit the potential capability to build mechanical tunable quantum NEMS such that the mechanical system can either suppress, enhance or even change the direction of electronic current.

\begin{figure}
\centering
\includegraphics[width=\columnwidth]{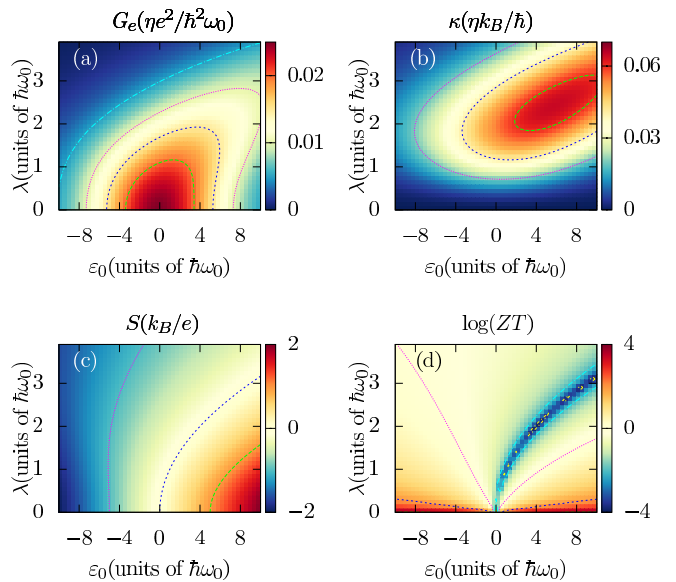}
\caption{\label{fig:zt}(Color online) Thermoelectric coefficients of the SET-NR system. Electronic conductance $G_e$ [panel (a)], thermal conductance $\kappa$ (panel b), Seebeck coefficient $S$ [panel (c)] and log of figure of merit $\mbox{log}(ZT)$ [panel (d)] are plotted as a function of the gate voltage parameter $\varepsilon_0$ and SET-NR coupling strength $\lambda$. The temperature $T=5\hbar\omega_0/k_B$ and chemical potential $\mu=0$. The coupling parameters and background temperature are the same as in Fig.~\ref{fig:current}.}
\end{figure}
Next we comprehensively study the thermoelectric transport properties of the SET-NR island in the linear response regime by plotting all the transport coefficients, namely the electronic conductance $G_e$, thermal conductance $\kappa$, Seebeck coefficient $S$, and figure of merit $ZT$, in Fig.~\ref{fig:zt}. Here we again clearly observe the polaron shift in the electronic conductance [Fig.~\ref{fig:zt}(a)] and Seebeck coefficient [Fig.~\ref{fig:zt}(c)] plots. The maximum electronic conductance [Fig.~\ref{fig:zt}(a))] and zero Seebeck coefficient [blue dashed line in Fig.~\ref{fig:zt}(c)] shift towards the righthand side for increasing $\lambda$. In the calculation of thermal conductance, since in our model the electrodes do not contain phonons, the entire contribution to the thermal conductance is from the electrons. We approximate the phonon contribution of the leads as a small constant added to the thermal conductance. This assumption is valid because the phonon contribution to thermal conductance is generally small in nano-junctions connected to metallic leads \cite{Zianni2010}. From Fig.~\ref{fig:zt}(b) we observe that in the regime of small $\lambda$ the thermal conductance approaches zero. This is because when the SET-NR coupling strength approaches zero, there is only one channel for the electron to tunnel through the SET island. In this case, since only the electron carries energy, the heat flow is only possible when there is some electron current. As a result, the thermal conductance is always zero.  Subsequently $ZT$ will be large due to the small thermal conductance. Therefore, materials with restricted tunneling channels, or delta-shaped transport distribution have been suggested as potentially good thermoelectric materials \cite{Mahan1996}. However, here we find that the SET-NR coupling can open up extra channels for tunneling so that the thermal conductance will increase quickly in the presence of a quantum NR. From Fig.~\ref{fig:zt}(d) we can see that $ZT$ decreases dramatically with the SET-NR coupling strength. This result is consistent with recent findings from different approaches such as the rate equation approach \cite{Zianni2010}, Green's function approach \cite{Koch2014} and semiclassical approaches \cite{Perroni2014}, where a decrease in $ZT$ is found by introducing vibrational coupling for most $\varepsilon_0$. 

\begin{figure}
\centering
\includegraphics[width=\columnwidth]{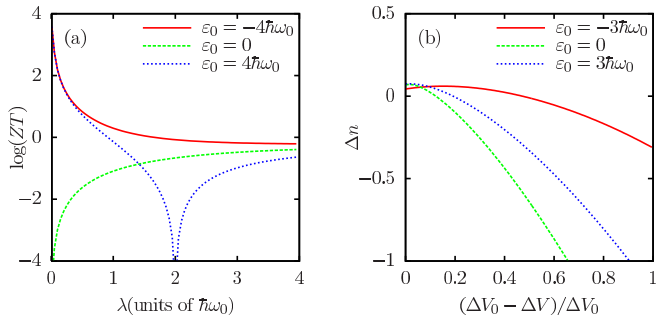}
\caption{\label{fig:line}(Color online) Panel a: SET-NR coupling strength dependence of $ZT$ for different gate voltage parameters $\varepsilon_0$. Panel b: Backaction ($\Delta n$) as a function of decreasing voltage bias and increasing temperature bias while keeping the electronic current constant. The paramters are $\lambda=\hbar\omega_0$, $\Delta V_0=4\hbar\omega_0/e$, and average $V$ at 0. In both panels, the coupling parameters and background temperature are same as Fig.~\ref{fig:current}.}
\end{figure}

Here we explicitly show the SET-NR coupling strength dependence in Fig.~\ref{fig:line}(a). We observe that $ZT$ decays rapidly (even in the log scale) for both positive and negative $\varepsilon_0$. This would cause a severe hindrance to experiments based on constrained tunneling to enhance $ZT$, since small (but finite) vibrational couplings are unavoidable in such systems. In this figure the position of the dip in the curve for $\varepsilon = 4\hbar\omega_0$ corresponds to the vanishing Seebeck coefficient $S$. In other words this is the position at which the energy spacing $\varepsilon_0$ is located exactly at the Fermi-level of the leads, taking the polaron shift into account. Thus the effect of a decreasing $\varepsilon_0$ is to shift the dip towards the left-hand side and eventually it vanishes for negative $\varepsilon_0$, in the process causing $\mbox{log}(ZT)$ to become negative for $\varepsilon_{0} = 0$. Even though at small values of SET-NR coupling strength $\lambda$, $\mbox{log}(ZT)$ rapidly decreases, it saturates for large values of $\lambda$ to $\approx ZT = 1$ and is no longer sensitive to the gate voltage. This saturation behavior is due to the fact that when the strength of the nonlinearity is quite high, the electronic and thermal conductance are always increasing or decreasing simultaneously as we can see from Figs.~\ref{fig:zt}(a) and \ref{fig:zt}(b). 

\subsection{Backaction on the NR}

The backaction of the electronic current on the NR states is of primary interest due to its effects on device sustainability or measurement sensitivity. The backaction can be strong in some cases \cite{Blanter2004,Clerk2005}, but for detector applications a small backaction is preferred to stabilize the NR leading to accurate measurements. It is therefore of immense interest to build devices in which the mechanical motion can tune the electronic properties effectively while the mechanical system itself is well sustained and not affected by the backaction of the electronic current. In order to study this backaction we investigate the effects on NR vibrational states due to the electronic current passing through the SET-NR island. If the NR is weakly coupled to the SET it will equilibrate to the background temperature. Therefore one would suspect that the expectation value of the energy level of the NR, $n_{eq}$, obeys the Bose-Einstein distribution $n_{eq}=[e^{\beta_{E} \hbar\omega_0}-1]^{-1}$ with the equilibrium temperature $k_B T_E = \beta^{-1}_{E}$. This assumption completely fails in the strong SET-NR coupling regime since the charges on the SET will affect the vibrational state of the NR. Hence we generalize the distribution of the NR to capture the strong SET-NR coupling effects. In order to do this we first obtain the canonical density matrix of the entire SET-NR island which equilibrates to the background environment given by $\rho \propto e^{-\beta_{E} H_{S}}$. Only after that we trace over the SET degrees of freedom to get the reduced density matrix of the NR. Using this approach, the distribution of the energy levels of the NR will no longer be a Bose-Einstein distribution but will be given \emph{exactly} by
\begin{equation}
\label{eq:polaron}
n_{eq}=\frac{1}{e^{\beta_{E}\hbar\omega_0}-1}+\frac{\lambda^2/(\hbar\omega_0)^2}{e^{\beta_{E}[\varepsilon_0-\lambda^2/(\hbar\omega_0)]}+1}.
\end{equation}
The derivation of the above equation can be found in Appendix \ref{sec:neq}. Clearly the first term accounts for the Bose-Einstein distribution, whereas the second term captures the effect of strong SET-NR coupling strength. This term can be understood physically as the formation of a polaron with energy $\lambda^2/(\hbar\omega_0)$. The polaron only exists when an extra charge is present on the island and the polaron energy serves as the chemical potential to that charge. Hence, on average the extra energy applied on the NR, due to the finite SET-NR coupling, follows a Fermi-Dirac distribution multiplied by the polaron energy as given by the second term in Eq.~(\ref{eq:polaron}). This second term can dominate over the first term in the strong SET-NR coupling regime or in the low temperature regime. In these regimes if the charging energy ($\varepsilon_0$) is smaller than the polaron energy then an increase in temperature would cause the average NR excitation number ($n_{eq}$) to decrease due to the reduced probability of the polaron formation.

In the nonequilibrium transport regime Eq.~(\ref{eq:polaron}) is no longer valid because the passage of electronic current will disturb the energy distribution of NR. However, the energy distribution of the NR can be obtained numerically with the help of the $0$th order reduced density matrix as, $n_{neq}=\mbox{Tr}(\rho^{(0)}a^\dagger a)$, where $a$ and $a^\dagger$ are creation and annihilation operators of the NR. This nonequilibrium energy distribution allows us to study the effects of the current on the NR energy distribution, i.e., the backaction, using a distribution difference defined as $\Delta n=n_{neq}-n_{eq}$.
\begin{figure}
\centering
\includegraphics[width=\columnwidth]{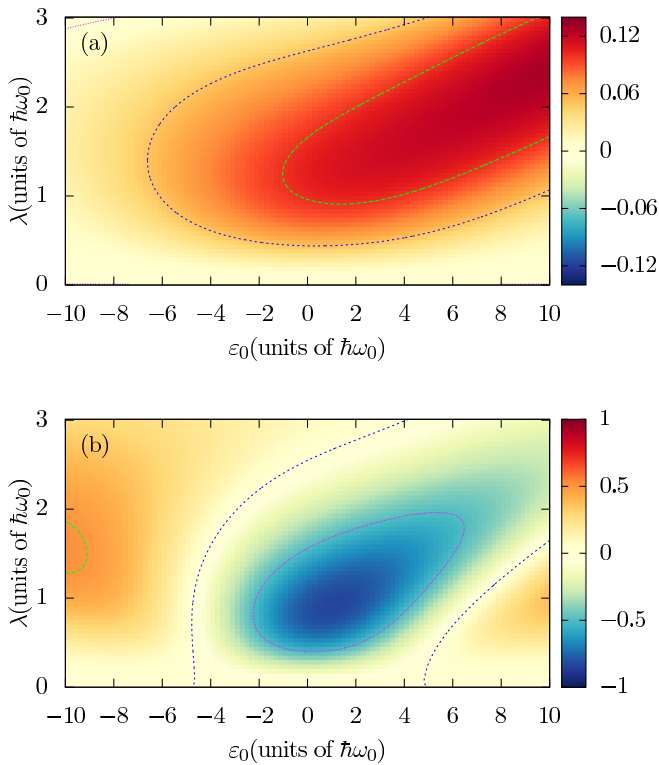}
\caption{\label{fig:dn}(Color online) Backaction ($\Delta n$) under voltage and temperature biased current. $\Delta n$ is plotted as a function of the gate voltage parameter $\varepsilon_0$ and SET-NR coupling strength $\lambda$. Panel (a) shows $\Delta n$ under voltage bias with $\mu_L=2\hbar\omega_0$ and $\mu_R=-2\hbar\omega_0$. Panel (b) shows $\Delta n$ under temperature bias $T_{L}=T_{E}+\Delta T$ and $T_{R}=T_{E}-\Delta T$ with $\Delta T=3\hbar\omega_0/k_B$. In both panels, the coupling parameters and background temperature are same as Fig.~\ref{fig:current}.}
\end{figure}
Figure~\ref{fig:dn} shows the contour plot of the backaction ($\Delta n$) as a function of $\varepsilon_0$ and $\lambda$ under both voltage and temperature bias. The red regions represent heating whereas the blue region represents cooling. The yellow regions represents the parameter regime where the backaction is not present as $\Delta n=0$. As we can see for small values of $\lambda$ the backaction almost vanishes in both cases, which is expected because in this situation the electronic system and the vibrational system are almost decoupled so the NR equilibrates to its background environment. Interesting effects appear in the intermediate to strong SET-NR coupling regime. Figure~\ref{fig:dn}(a) shows $\Delta n$ under the voltage bias condition and we find in this case $\Delta n$ is always positive. In principle voltage induced cooling can be found in more complicated systems such as a double quantum dot\cite{Zippilli2010} or a superconducting SET\cite{Naik2006}. However, for the simple setup in our work, which can be experimentally easily realized, one always expects heating in the weak system-lead coupling limit ($\eta\ll\omega_0$)\cite{Piovano2011}. However, for temperature bias situation as shown in Fig.~\ref{fig:dn}(b) we find both positive and negative $\Delta n$, which indicates that it is even possible to cool the NR. Importantly, we find that for every $\lambda$ there are two corresponding values of $\varepsilon_0$ where $\Delta n$ vanishes [dashed blue lines in Fig.~\ref{fig:dn}(b)], which indicates that backaction can be eliminated by external adjustment of the gate voltage. Furthermore, these two values are located exactly where the thermoelectric current reaches maximum (see Fig.~\ref{fig:current}). This clearly implies that the temperature biased current is able to achieve the primary goal of producing current effectively without severely affecting the NR vibrational state. In Fig.~\ref{fig:line}(b) we show the backaction on the NR, characterized by $\Delta n$, caused by a fixed amount of electronic current with different weights of voltage and temperature bias. If we reduce the voltage bias from an initial bias ($\Delta V_0$) to a smaller voltage bias ($\Delta V$) while adding temperature bias properly to keep the electronic current constant, we find that $\Delta n$ goes from positive to zero and eventually negative. By adding a temperature bias, it is thus possible to find a suitable gate voltage such that $\Delta n=0$. Thus, temperature biased currents could be a possible solution for SET-NR detector applications to keep the backaction to a bare minimum.

\section{\label{summary}Summary}
We have investigated the thermoelectric transport properties of a SET coupled to a quantum NR and importantly its dependence on the SET-NR coupling strength. We demonstrated that a quantum NR is capable of suppressing, enhancing or even changing the direction of thermoelectric current of the SET. This is because the NR and the electron form a polaron which can effectively shift the charging energy of the SET. The charging energy controls the dominant carrier type (holes or electrons) and thus leads to a change in the direction of thermoelectric current.

Furthermore, we have shown that even a small SET-NR coupling can dramatically suppress $ZT$ because a finite SET-NR coupling strength will introduce extra tunneling channels for electrons which will greatly enhance the thermal conductance and thus reduce $ZT$. On the other hand, in the strong SET-NR coupling regime we find that $ZT$ saturates and becomes insensitive to the gate voltage. 

The backaction of the electronic current on the NR is also examined and we have observed that cooling of the NR is possible for electronic current under a temperature bias (thermoelectric current), whereas in the standard voltage bias regime only heating of the NR is possible. Importantly, we also find that it is possible to eliminate the backaction in the parameter regimes where one can effectively generate the thermoelectric current. As a result, we propose that thermoelectric currents can be a possible solution for detector applications, where one needs an electronic current passing through the SET without immensely affecting the NR.

\appendix
\section{Correlation functions}
\label{sec:correlation}
In this appendix we give explicit formulas for the correlation functions used in our calculations. For the spectral densities chosen in this work, all the correlation functions can be explicitly evaluated in terms of the Matsubara summations, by using the residue theorem. Here we provide the results. 

The correlation functions used in the quantum master equation are given by
\begin{eqnarray}
C^{12}(t)&=&\mathcal{F}_L(t)+\mathcal{F}_R(t),\\
C^{21}(t)&=&\frac{(\eta_L+\eta_R)\varepsilon_D}{2}e^{-\varepsilon_Dt/\hbar}-C^{12}(-t).
\end{eqnarray}
When $t>0$, the function $\mathcal{F}_\alpha(t)$ reads
\begin{eqnarray}
\mathcal{F}_\alpha(t)&=&\sum_{l=1,3,5,...}\Bigl[\frac{1}{\beta_\alpha}\frac{\mathrm{i}\eta_\alpha\varepsilon_D^2}{(\mu_\alpha-\mathrm{i}v_l^{\alpha})^2+\varepsilon_D^2}e^{-(v_l^{\alpha}+\mathrm{i}\mu_\alpha)t/\hbar}\Bigr]\nonumber\\
&&+\frac{\eta_\alpha\varepsilon_D}{2[e^{-\beta_\alpha(\mu_\alpha+\mathrm{i}\varepsilon_D)}+1]}e^{-\varepsilon_Dt/\hbar}.
\end{eqnarray}
 In case of $t<0$ the function $\mathcal{F_\alpha}$ can be obtained using $\mathcal{F}_\alpha(t)=\mathcal{F}_\alpha^*(-t)$. Here $v_l^{\alpha}=\pi l/\beta_{\alpha}$ is the Matsubara frequency. 

The correlation functions used in the current operators are
\begin{eqnarray}
\mathcal{C}_e^{12}(t)&=&-e\mathcal{F}_L(t),\\
\mathcal{C}_e^{21}(t)&=&\frac{e\eta_L\varepsilon_D}{2}e^{-\varepsilon_Dt/\hbar}+\mathcal{C}_e^{12}(-t),\\
\mathcal{C}_h^{12}(t)&=&-\mathcal{H}_L(t),\\
\mathcal{C}_h^{21}(t)&=&\frac{\eta_L\varepsilon_D(\varepsilon_D-\mu_L)}{2}e^{-\varepsilon_Dt/\hbar}+\mathcal{C}_h^{12}(-t),
\end{eqnarray}
where
\begin{eqnarray}
\mathcal{H}_\alpha(t)&=&\sum_{l=1,3,5,...}\Bigl[\frac{1}{\beta_\alpha}\frac{\mathrm{i}\eta_\alpha (v_l^{\alpha}-\mu_\alpha)\varepsilon_D^2}{(\mu_\alpha-\mathrm{i}v_l^{\alpha})^2+\varepsilon_D^2}e^{-(v_l^{\alpha}+\mathrm{i}\mu_\alpha)t/\hbar}\Bigr]\nonumber\\
&&+\frac{\eta_\alpha\varepsilon_D(\varepsilon_D-\mu_\alpha)}{2\left[e^{-\beta_\alpha(\mu_\alpha+\mathrm{i}\varepsilon_D)}+1\right]}e^{-\varepsilon_Dt/\hbar},
\end{eqnarray}
for $t>0$. Similarly we can use the relation $\mathcal{H}_\alpha(t)=\mathcal{H}_\alpha^*(-t)$ to obtain $\mathcal{H}_\alpha(t)$ when $t<0$. 

\section{Equilibrium phonon distribution}
\label{sec:neq}
When the SET-NR system is in thermal equilibrium with its environment, the reduced density matrix will follow the canonical distribution $\rho=e^{-\beta_EH_S}/Z_S$, where $Z_S=\mathrm{Tr}(e^{-\beta_EH_S})$. Therefore, the average excitation number is given by
\begin{equation}
n_{eq}=\mathrm{Tr}\left(e^{-\beta_E H_S}a^\dagger a\right)/Z_S.
\end{equation}
The explicit form of the above equation can be evaluated with the help of polaron transformation $\bar{O}=e^SOe^{-S}$ where $S=\frac{\lambda}{(\hbar\omega_0)}d^\dagger d(a^\dagger-a)$ and $O$ is an arbitrary operator. One can then explicitly find the operators in the polaronic frame $\bar{a}=a-\lambda/(\hbar\omega_0)d^\dagger d$, $\bar{a}^\dagger=a^\dagger-\lambda/(\hbar\omega_0)d^\dagger d$ and $\bar{\rho}=e^{-\beta_E\bar{H}_S}/Z_S$ with
\begin{equation}
\bar{H}_S=(\varepsilon_0-\frac{\lambda^2}{\hbar\omega_0})d^\dagger d+\hbar\omega_0a^\dagger a.
\end{equation}
Therefore, the equilibrium distribution of the phonon is given by 
\begin{eqnarray}
n_{eq}&=&\mathrm{Tr}\left(\bar{\rho}\bar{a}^\dagger \bar{a}\right)\nonumber\\
&=&\mathrm{Tr}\left[\bar{\rho}\Big(a^\dagger a-\frac{\lambda}{\omega_0}d^\dagger d(a^\dagger+a)+\frac{\lambda^2}{\omega_0^2}d^\dagger d\Big)\right]\label{eq:polaron1}\\
&=&\frac{1}{e^{\beta_{E}\hbar\omega_0}-1}+\frac{\lambda^2/(\hbar\omega_0)^2}{e^{\beta_{E}[\varepsilon_0-\lambda^2/(\hbar\omega_0)]}+1},\label{eq:polaron2}
\end{eqnarray}
where the first term in Eq. (\ref{eq:polaron2}) denotes the Bose-Einstein distribution of the NR mode and the second term in Eq. (\ref{eq:polaron2}) denotes the contribution from the polaron.

\section*{Acknowledgements}
This work is supported by Ministry of Education (MOE), Singapore by  Grant No. MOE 2012-T2-1-114.
\bibliography{qme-ss}
\end{document}